\documentclass[%
 reprint,
 superscriptaddress,
 amsmath,amssymb,
 aps,
 prl,
]{revtex4-1}

\usepackage{graphicx}
\usepackage{dcolumn}
\usepackage{bm}
\usepackage{hyperref}

\hypersetup{
	colorlinks = true,
	pdftitle = {},
	pdfauthor = {},
	pdfkeywords = {},
	linkcolor = blue,
	citecolor = blue,
	filecolor = black,
	urlcolor = magenta
}

\begin{document}


\title{Proposal for a three-dimensional magnetic measurement method with nanometer-scale depth resolution}%

\author{Devendra Negi}
\affiliation{Department of Physics and Astronomy, Uppsala University, P.O. Box 516, 75120 Uppsala, Sweden}%
\author{Lewys Jones}%
\affiliation{Advanced Microscopy Laboratory, Centre for Research on Adaptive Nanostructures and Nanodevices (CRANN), Dublin 2, Ireland}
\affiliation{School of Physics, Trinity College Dublin, Dublin 2, Ireland}
\author{Juan-Carlos Idrobo}%
\affiliation{Center of Nanophase Materials Sciences, Oak Ridge National Laboratory, Oak Ridge, TN 37831, USA}
\author{J\'{a}n Rusz}%
\email{jan.rusz@physics.uu.se}
\affiliation{Department of Physics and Astronomy, Uppsala University, P.O. Box 516, 75120 Uppsala, Sweden}%

\date{\today}

\begin{abstract}
We propose a magnetic measurement method based on combining depth sectioning and electron magnetic circular dichroism in scanning transmission electron microscopy. Electron vortex beams with large convergence angles, as those achievable in current state-of-the-art aberration correctors, could produce atomic lateral resolution and depth resolution below 2~nm.
\end{abstract}


\maketitle

Progress in the applications and development of magnetic nanostructures calls for the development of measurement methods capable of providing information at sufficiently high spatial resolution. Existing methods such as spin-polarized scanning tunneling microscopy \cite{spstm,spstmafm}, magnetic exchange force microscopy \cite{magefm}, x-ray magnetic circular dichroism (XMCD; \cite{xmcd15,xmcdptycho}) or electron holography \cite{magholo,holomag}, lack either spatial resolution or depth sensitivity. Recently, combining XMCD with tomography has allowed experimentalists to map {the average directions of magnetic moments} with a spatial resolution of about 100~nm \cite{xmcdtom}.

Transmission electron microscopy (TEM) offers high resolution, routinely reaching atomic resolution in today's aberration corrected instruments. It also allows magnetic studies via the method of electron magnetic circular dichroism (EMCD, \cite{nature}) at high lateral resolution. Recent EMCD studies done with atomic size electron beams, in scanning (S)TEM have succeeded in extracting magnetic information from sample areas of few square nanometers \cite{c34exp,apremcd,largealpha}. An alternative setup based on high-resolution TEM imaging has allowed for the detection of quantitative magnetic information from individual atomic planes \cite{elspemcd}. Yet, in both cases depth information is so far missing, with the observed data being a two-dimensional projection of the three-dimensional sample.

Depth-sectioning in the STEM was first implemented in the high-angle annular dark field mode (HAADF), detecting electrons scattered (quasi-)elastically to large angles, typically between 80--200~mrad. This has been used to detect the three-dimensional position of dopant atoms \cite{ds3datom,ds3dhaf,ds3dpnas,ds3dum,ds3dzeol,ds3dxin}, or the inclination of dislocations \cite{dsdisgan,dsdisscrew}. Recent HAADF simulations with large convergence angles have also shown how depth-sectioning could be used to measure sample thickness, study surface reconstruction or detect impurity atoms \cite{broek,ishikawa}. Going beyond HAADF, theoretical prediction \cite{dseelstheo} and later experiments by Pennycook et al.\ \cite{dseelsexp} have demonstrated nanometer scale elemental mapping combining depth sectioning atomic-resolution STEM with electron energy-loss spectroscopy (EELS).

In this Letter we introduce depth sectioning to the domain of magnetic studies. We propose to use electron vortex beams (EVBs; \cite{vorttem,mcmorran,vortjo}) of atomic size \cite{darius,atvort} to scan over a region of sample. The magnetic signal (EMCD) is extracted as the difference between two spectra, one measured with a beam carrying orbital angular momentum (OAM) $+\hbar$ and another with $-\hbar$. As we demonstrate using a simulated experiment, depth sensitivity is achieved by sweeping the focal plane through the thickness of the sample.

\begin{figure}[t]
  \includegraphics[width=7.5cm]{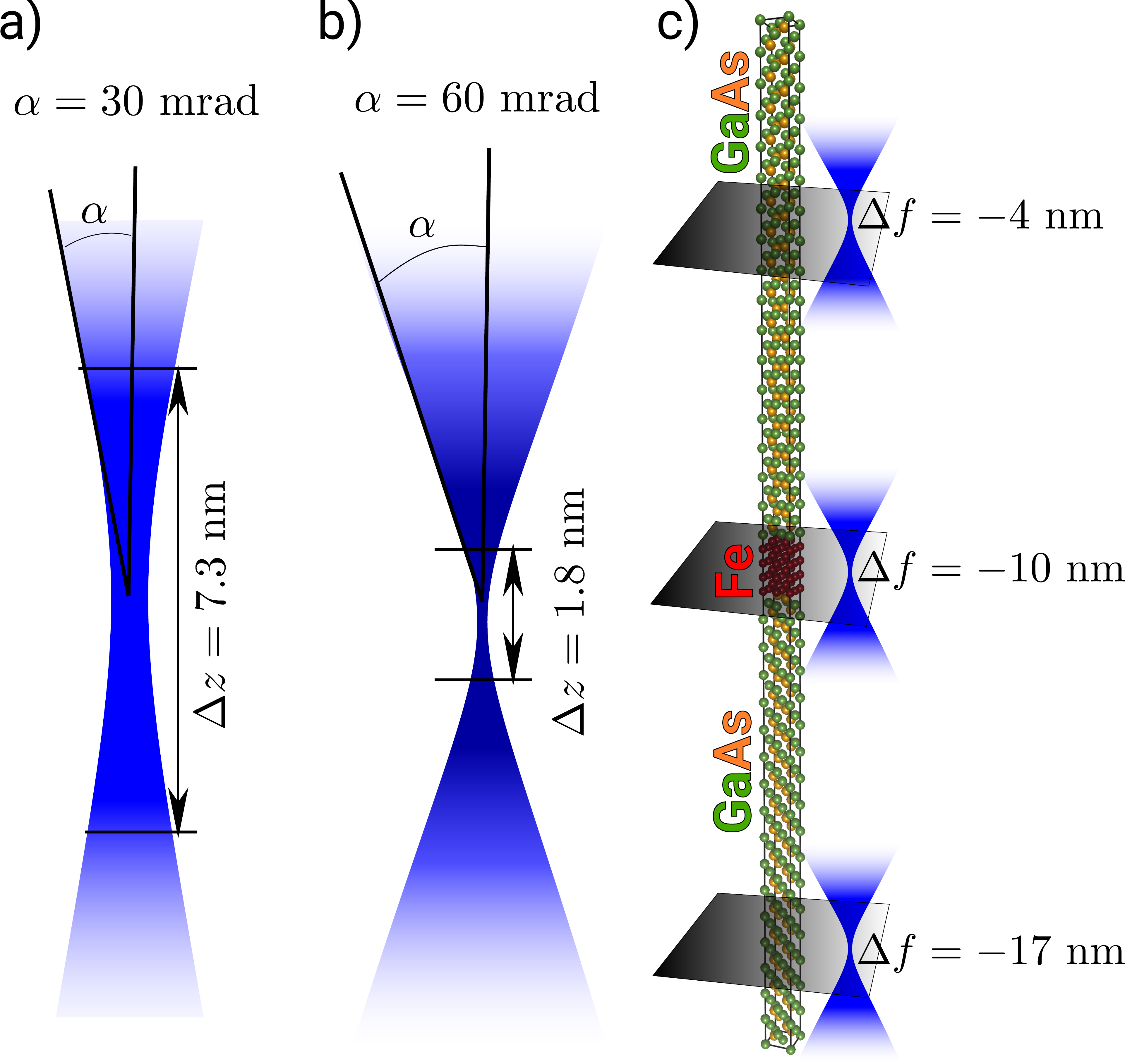}
  \caption{Schematic drawing of a convergent electron beam with convergence semi-angle a) $\alpha=30$~mrad and b) $\alpha=60$~mrad. The depth of focus parameter $\Delta z$ is indicated. c) Structure model of a sample studied in this Letter: 1~nm layer of bcc iron sandwitched between 10~nm thick layers of GaAs. Three focal planes are illustrated with their associated defocus parameter $\Delta f$ (see text for details).}
  \label{fig:scheme}
\end{figure}

First we describe the proposed experimental setup. In STEM, an image is created by focusing a convergent probe that is scanned over a region of a sample.  The intensity in every pixel of the image is the result of the integrated intensity of the electrons hitting the detector for each electron probe position.  Minimal diameter of the electron beam, and thus the the best spatial resolution of the image, is determined mostly by the convergence semi-angle $\alpha$ and by geometric and chromatic aberrations of the probe shaping electronics. Progresses in aberration corrections allowed to increase $\alpha$, by which the probe sizes less than $1$~\AA{} can be routinely achieved. Increasing $\alpha$ influences also another important electron beam parameter, namely, the depth of focus, $\Delta z$. Depth of focus describes the range of the $z$-coordinates around the focal plane, within which the probe diameter increases less than by a factor of $\sqrt{2}$. Depth of focus is inversely proportional to {$\alpha^2$}, given approximately by relation $1.77\lambda/\alpha^2$ \cite{nellistdof} where $\lambda$ is the de Broglie wavelength of beam electrons. Doubling the convergence angle reduces depth of focus four times. See Fig.~\ref{fig:scheme}a,b, where two electron beams with different convergence semi-angles, $\alpha=30$ and $\alpha=60$~mrad, are schematically illustrated. By shifting the focal plane up or down one can focus the electron beam into various depths of the sample.  This is readily achieved at the microscope by adjusting the defocus setting, $\Delta f$, as is illustrated in Fig.~\ref{fig:scheme}c. Here we assume the convention $\Delta f=0$ if the focal plane is at the entrance surface of the sample. $\Delta f$ is defined as negative (underfocused beam) when the focal plane is inside of the sample.

\begin{figure}[t]
  \includegraphics[width=8.6cm]{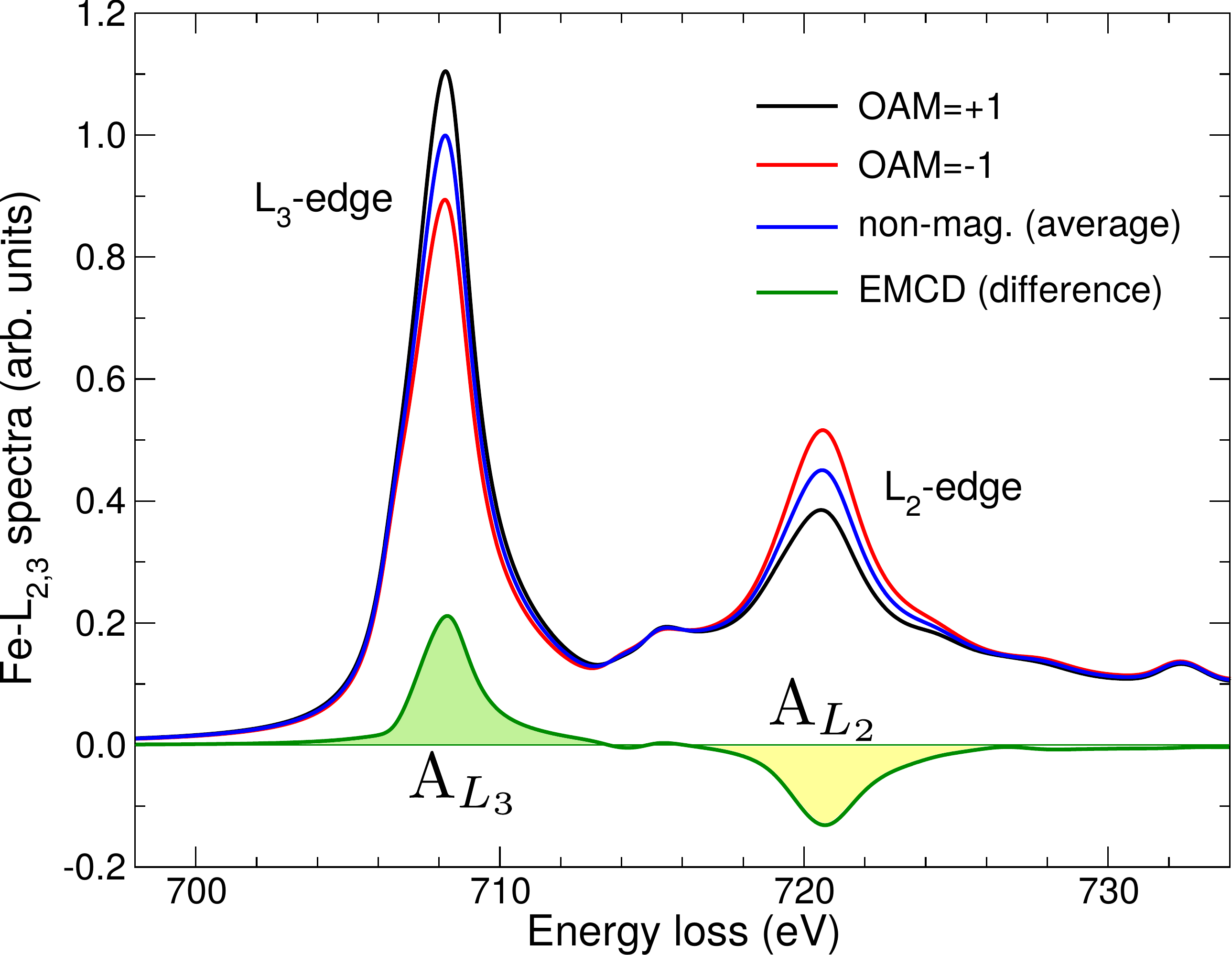}
  \caption{Density functional theory calculations of $L_{2,3}$-edge spectra of bcc iron. Example spectra calculated with OAM of $\pm \hbar$, their difference (EMCD) and average are being shown. Areas under the $L_3$ and $L_2$ edge of the EMCD spectrum, $A_{L_3}$ and $A_{L_2}$, can be used to determine magnetic properties via sum rules \cite{oursr,lionelsr}.}
  \label{fig:spectra}
\end{figure}

The electron beam scatters when interacting with the sample. Inelastic scattering processes cause beam electrons to lose part of their kinetic energy to various excitation processes in the sample.  Excitations of core level electrons in the sample lead to characteristic near-edge structures in the energy spectrum of scattered electrons.  Like in x-ray absorption spectroscopy, they give a wealth of information about the material's chemistry and electronic structure, as well as magnetism. In analogy with XMCD, electrons can be used to detect magnetic information as well, via the EMCD method. In an EMCD experiment \cite{nature}, two electron energy-loss spectra are being acquired under specific conditions and their difference, the EMCD spectrum (see Fig.~\ref{fig:spectra}), carries information about the magnetic properties of the sample. Specifically the areas within the EMCD spectrum at $L_3$ and $L_2$ peak edges, respectively, are used to obtain the ratio between the spin and orbital magnetic moments of the sample using sum rule expressions \cite{oursr,lionelsr}.

\begin{table}[ht]
  \caption{Tabulated depth-of-focus, $\Delta z$, as a function of beam convergence semi-angle, $\alpha$, and acceleration voltage.}
  \label{tab:dof}
  \begin{tabular}{lcccc}
    \hline \hline
                      & \multicolumn{4}{c}{Acceleration voltage} \\
    $\Delta z$ (nm)   & 80~kV & 100~kV & 200~kV & 300~kV \\
    \hline
    $\alpha= 8$~mrad  &  116  &  102  &  69  &  54\cite{largealpha} \\
    $\alpha=12$~mrad  &   51  &   46  &  31  &  24  \\
    $\alpha=15$~mrad  &   33  &   29  &  20\cite{apremcd}  &  16  \\
    $\alpha=20$~mrad  &   19  &   16  &  11  &  8.7 \\
    $\alpha=30$~mrad  &   8.2 &   7.3\cite{c34exp} &  4.9 &  3.9 \\
    $\alpha=60$~mrad  &   2.1 &   1.8 &  1.2 &  1.0 \\
    $\alpha=100$~mrad &   0.7 &   0.7 &  0.4 &  0.3 \\
    \hline \hline
  \end{tabular}
\end{table}

Until recently, EMCD measurements were performed with relatively small convergence angles, $\alpha$. Therefore the depth of focus, $\Delta z$, exceeded typical sample thicknesses. This remains true even for very recent STEM-EMCD experiments with atomic size aberrated electron beams. At $\alpha=8$~mrad and an acceleration voltage of 300~kV, used in Ref.~\onlinecite{largealpha}, the depth of focus is approximately 54~nm. In another work, where an EMCD measurement method with atomic plane resolution was presented \cite{apremcd}, $\alpha=15$~mrad and acceleration voltage 200~kV leads to $\Delta z = 20$~nm. However, opening the convergence angle further to $\alpha=30$~mrad, the $\Delta z$ at 100~kV or 200~kV becomes 7.3~nm or 4.9~nm, respectively, which is less than the typical TEM sample thickness. EMCD experiments with such settings have been performed at $\alpha=30$~mrad and $V_{acc}=100$~kV using four-fold astigmatic probes \cite{c34exp}, which have however different focusing properties despite being of atomic size. For convenience, Tab.~\ref{tab:dof} summarizes depths of foci as a function of acceleration voltage and convergence semi-angle.

In 2010 it was proposed that EVBs can be used as efficient probes for magnetic measurements \cite{vortjo}. Later theoretical considerations refined this picture by narrowing it down to the atomic resolution domain only \cite{vortexelnes,schattnp}. Atomic size electron vortex beams have since been demonstrated \cite{darius,atvort} and other methods of their preparation have been described \cite{krivanek}.

Combining the use of EVBs of atomic size with depth sectioning we gain access magnetic information in all three dimensions. At 100~kV and convergence semi-angles of $\alpha=30$~mrad and $\alpha=60$~mrad we obtain depths of foci of 7.3~nm and 1.8~nm, respectively. While aberration corrected beams with a convergence semi-angle of 30~mrad are a routine task for today's modern aberration-corrected STEM, a corrected probe with 60~mrad convergence semi-angle is now a reality with the latest state-of-the-art of aberration correctors \cite{dellby,sawada,hosokawa}. Yet, technological progress is bound to enable such experiments in a wider range of instruments within a horizon of a few years.

To verify our predictions of the use of EMCD depth sectioning, a computational experiment has been designed to reflect realistic experimental conditions. A structure model consisting of four unit cells of bcc iron (approximately 1.1~nm thick) is sandwiched between two 9.6~nm thick layers of GaAs \cite{gaasfe}. The constructed unit cell, tiled periodically in the $x$ and $y$-directions, is visualized in Fig.~\ref{fig:scheme}c. We utilize the combined multislice / Bloch waves method for simulations of inelastic electron scattering, as implemented in software \textsc{mats.v2} \cite{matsv2}. Electron vortex beam wavefunction at the entrance surface of the sample was generated as a Fourier transform of 
$$\psi(k,\phi)=e^{im\phi} e^{-\pi \lambda \Delta f k^2} \Theta(2\pi\alpha/\lambda-k),$$
where $m=\pm 1$ for OAM of $\pm \hbar$, $\Theta(x)$ is Heaviside step function, and $k,\phi$ are the cylindrical coordinates in reciprocal space. EVB was centered on an atomic column, collection semi-angle was set to $10$~mrad and defocus was varied from zero to $-20$~nm.

\begin{figure}[t]
  \includegraphics[width=8.6cm]{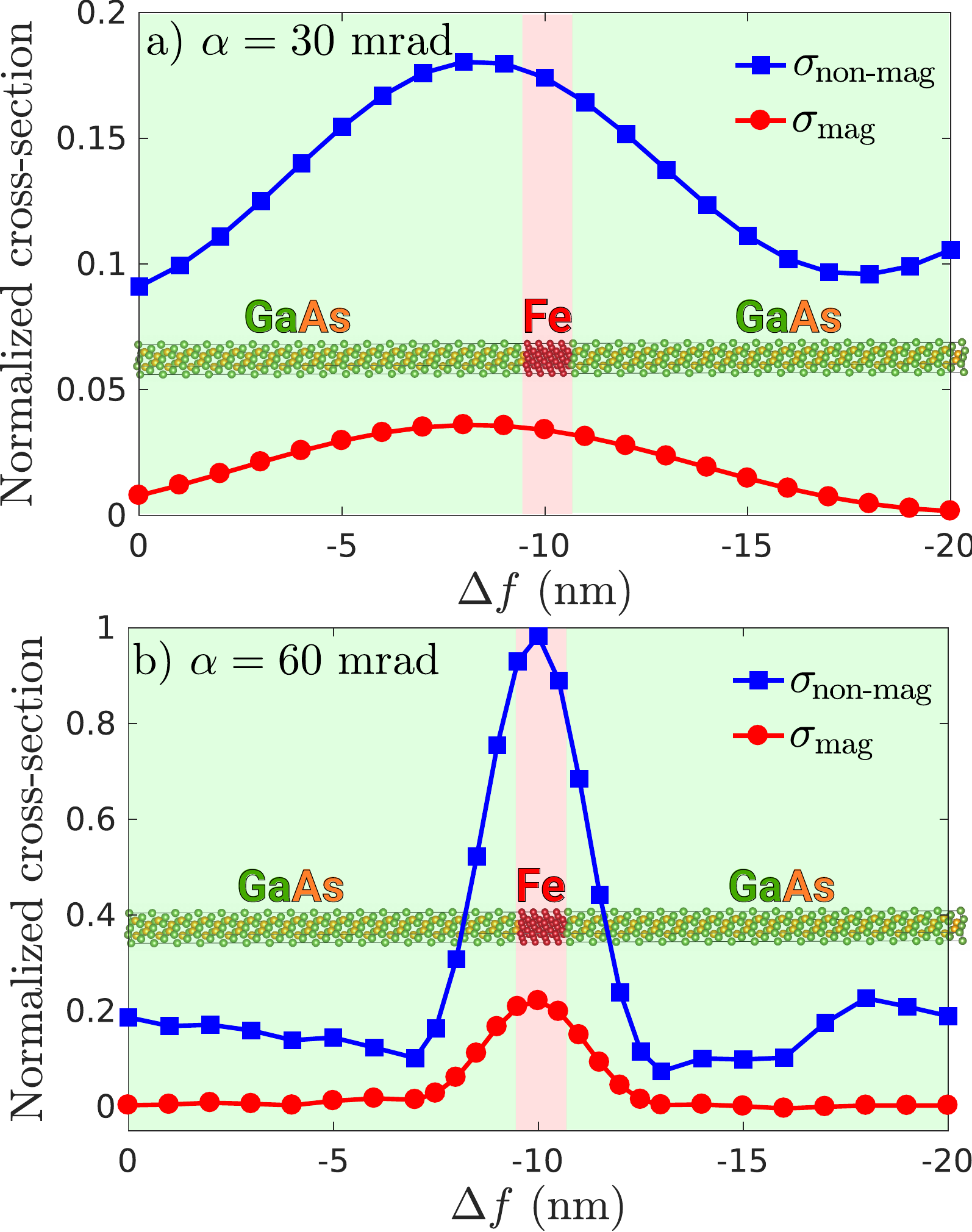}
  \caption{Simulated magnetic depth sectioning experiment on the GaAs/Fe/GaAs structure model (total sample thickness 20.3~nm) at convergence angle a) $\alpha=30$~mrad{, b) and c)} $\alpha=60$~mrad, respectively. {Panels b) and c) differ in the position of magnetic layer, as indicated by the inset structure model. Acceleration voltage was set to 100~kV and beam direction along (001) zone axis.} Blue line shows $L_3$ edge integral of averaged spectrum (non-magnetic part of the inelastic scattering cross-section, $\sigma_\text{non-mag}$) and red line shows $L_3$ edge integral of difference spectrum (magnetic part, $\sigma_\mathrm{mag}$, i.e., EMCD).}
  \label{fig:result}
\end{figure}

Figure~\ref{fig:result} shows energy integrals of the average electron energy-loss (EEL) spectrum ($\sigma_\text{non-mag}$) and the (EMCD) spectrum ($\sigma_\mathrm{mag}$) for the $L_3$ edge, up to 718~eV energy loss (see Fig.~\ref{fig:spectra}), as a function of defocus $\Delta f$. Calculations at both convergence angles show a peak nearby the position of the iron layer. At 30~mrad convergence angle, the maxima of the integrated EEL and EMCD spectra are shifted about 2~nm before the actual iron layer begins. A possible reason is the strong channeling of EVBs through an atomic column \cite{loffler,xin,mendis,nguyen}, which onsets once the vortex diameter reaches a sufficiently low value. In fact, a much smaller yet noticeable shift of the peak magnetic cross-section can also be observed at 60~mrad convergence angle. However, with increased convergence angle, both precision (depth-of-focus) and accuracy (position of the peak relative to the position of iron layer) clearly improves. {We have verified that the depth-effect is not related to the \emph{Pendell\"{o}sung} oscillations, which are commonly observed in EMCD studies \cite{prbtheory,lofflercode}. Placing the magnetic layer to another depth within the multilayer model also shifts the peaks of non-magnetic and magnetic components of inelastic scattering cross-sections as a function of defocus (not shown).}

One important result of the calculations is that the EMCD signal converges to zero values, once the focal plane is far from the magnetic layer. This contrasts with the average (non-magnetic) cross-section, which is always nonzero. This is likely due to the requirement of atomic size vortex structures around atomic columns in order to detect an EMCD signal. In contrast, the non-magnetic component of the scattering cross-section to be nonzero only needs the beam electrons to be sufficiently close to the iron atoms, which they always do since the iron layer is extended in the $x,y$-plane. This suggests that depth sectioning should provide improved selectivity in the  magnetic cross section when compared with the non-magnetic counterpart \cite{dseelstheo,dseelsexp}.

Having demonstrated from the simulation study the potential of this method, we now consider the practical advice to successfully realise such a measurement which will depend on the attainable signal-to-noise-ratio (SNR). Here, by signal we understand the energy-integral of the EMCD spectrum, defined as the area $A_{L_3}$ in Fig.~\ref{fig:spectra}, and the noise is determined by the energy-integral of the average spectrum, $B_{L_3}$ plus a background signal, which was not simulated in Fig.~\ref{fig:spectra}, but is inevitably present in experiments.  Here we make an assumption that the background signal contributes about $b=2$ times the average edge signal. Assuming purely Poissonian noise, per spectrum the signal to noise ratio is then $A_{L_3}/\sqrt{{2}(1+b) B_{L_3}}$. Since in our simulations \cite{fnote1} $A_{L_3}\approx 0.12 B_{L_3}$, we obtain SNR of ${0.085} \sqrt{B_{L_3}/(1+b)}$.

In a typical atomic resolution spectrum imaging experiment the electron beam scans over the chosen region of the sample, collecting thousands of spectra.
{Modern experimental practice for EELS mapping allows to sample atomic columns by finely spaced probe positions. Typically each atomic column is sampled by a small cloud of pixels. From previous numerical studies we know that EMCD decreases quickly with increasing distance of the EVB from the atomic column \cite{vortexelnes}, therefore as a safe compromise for accumulation of a magnetic signal from spectrum image, one should be able to integrate over a} $N_\mathrm{pix}=3 \times 3$ pixel region centered about a single atomic column position.
Then the final SNR becomes ${0.085} \sqrt{N_\mathrm{pix} B_{L_3}/(1+b)} = {0.15}\sqrt{B_{L_3}}$ for parameter values assumed here. In an experiment, where one just aims for detection of the magnetic layer, a requirement of $\text{SNR}>3$ leads to $B_{L_3}>{400}$ counts, {i.e., within an order of few hundreds of counts. Advances in instrument stability and new best-practices including fast multi-frame spectroscopy \cite{lewys} make this increasingly achievable with todays technology.}. For quantitative measurements a much higher SNR will be needed and the required counts will increase accordingly.

{



Technology today is on a verge of realizing both the precursor experiments and their combination should be the next step. First, aberration correctors offering corrected probes with well over 60~mrad convergence semi-angles already exist \cite{dellby,sawada,hosokawa} and it is only a question of time, when they will be applied to depth-sectioning experiments. Second, as of today, atomic size EVBs have been produced \cite{darius,atvort} and EMCD experiments with them are likely in progress. Third, progresses in data processing techniques have enabled to collect multiple spectrum images at reduced doses, to be stacked afterward \cite{lewys}. Using this method it is possible to counter scan noise and significantly reduce the dwell times. Simultaneously, modern denoising strategies utilizing multi-dimensional character of the spectrum images \cite{llr} provide efficient means to reduce the influence of noise thanks to considering local correlations in both spatial and energy dimensions. Acquiring focal series of spectrum images provides an additional dimension to the dataset and exploiting the correlation of spectrum images as a function of defocus will enable detection of magnetic layers at lower count rates than is required for a single spectrum image.
}

In summary, we have proposed a three-dimensional magnetic measurement method which can offer atomic lateral resolution and few-nm depth resolution. This is achieved by utilizing spectrum images collected with electron vortex beams at various defoci. We have estimated the feasibility of such magnetic measurements from a realistic size of specimen and at realistic signal collection parameters, giving guidance to future experimentalists about the data quality required for detection of an EMCD signal with this nanometer scale depth resolution and atomic lateral resolution. {We take this Letter as an opportunity to think ahead of the future applications of EVBs so that we can inform and engage in the process of developing the targets for the instrumentation to achieve. Delivering a technology enabling to measure magnetization at nanometer scale volume resolution should provide such incentive.} Successful experimental realization of our proposal will play a crucial role in design of magnetic nano-structures across a whole range of today's and tomorrow's electronic devices.

DSN and JR acknowledge funding from Swedish Research Council, G\"{o}ran Gustafsson's foundation and Carl Tryggers foundation.  JCI acknowledges support from the Center for Nanophase Materials Sciences, which is a DOE Office of Science User Facility.  Calculations were performed using resources of the Swedish National Infrastructure for Computing at NSC Center.


\begin{thebibliography}{99}
\bibitem{fnote1} {Note that simulations allow to cleanly separate the $L_3$ and $L_2$ edges and the reported $A_{L_3}/B_{L_3}$ ratio assumed integration over the whole energy range for the $L_3$ edge.}
\end{thebibliography}
\end{document}